\documentclass[]{aa}  

\makeatletter
\renewcommand*\aa@pageof{, page \thepage{} of \pageref*{LastPage}}
\makeatother
\usepackage{graphicx}
\usepackage{txfonts}
\usepackage[]{hyperref}
\usepackage{listings}
\usepackage{xcolor}
\definecolor{codegreen}{rgb}{0,0.6,0}
\definecolor{codegray}{rgb}{0.5,0.5,0.5}
\definecolor{codepurple}{rgb}{0.58,0,0.82}
\definecolor{backcolour}{rgb}{0.95,0.95,0.92}

\lstdefinestyle{mystyle}{
  backgroundcolor=\color{backcolour}, commentstyle=\color{codegreen},
  keywordstyle=\color{magenta},
  numberstyle=\tiny\color{codegray},
  stringstyle=\color{codepurple},
  basicstyle=\ttfamily\footnotesize,
  breakatwhitespace=false,         
  breaklines=true,                 
  captionpos=b,                    
  keepspaces=true,                 
  numbers=left,                    
  numbersep=5pt,                  
  showspaces=false,                
  showstringspaces=false,
  showtabs=false,                  
  tabsize=2
}

\begin{document}

   \title{
Adaptive Data Reduction Workflows for
Astronomy - 
   The ESO Data Processing System (EDPS)   
   }

   \titlerunning{The ESO Data Processing System (EDPS)}
   \authorrunning{Freudling et al.}

   \author{W. Freudling,
          S. Zampieri, 
          L. Coccato,
          S. Podgorski,
          M. Romaniello,
          A. Modigliani,           \and
          J. Pritchard 
          }

   \institute{European Southern Observatory, Karl-Schwarzschild-Str. 2, 85748 Garching, Germany,
              \email{wfreudli@eso.org}
             }

   \date{Received August 3, 2023; accepted November 3, 2023}


\abstract
{Astronomical data reduction is usually done with processing pipelines that
consist of a series of individual processing steps that can be executed
stand-alone. These processing steps are then strung together into workflows and
fed with data to address a particular processing goal.  Examples of such
pipeline processing goals are quality control of incoming data from telescopes,
unsupervised production of science and calibration products for an archive, and
supervised data reduction to serve the specific science goals of a scientist.
{For each of these goals, individual workflows need to be developed.  These
workflows need to evolve when the pipeline, observing strategies or calibration
plans change.  Writing and maintaining such a collection of workflows is
therefore a complex and expensive task.} }
{ In this paper, we propose a data processing system that automatically derives
processing workflows for different use cases from a single specification of a
cascade of processing steps. }
{ The system works by using formalized descriptions of data processing
pipelines that specify the input and output of each processing step. Inputs can
be existing data or the output of a previous step. Rules to select the most
appropriate input data are directly attached to the description. }
{ A version of the  proposed system has been implemented as the {\em ESO Data
Processing System (EDPS)} in the Python language. The specification of
processing cascades and data organisation rules use a restrictive set of Python
classes, attributes and functions.  } 
{The EDPS implementation of  the proposed system was used to demonstrate that
it is possible to automatically derive from a single specification of a
pipeline processing cascade the workflows that the European Southern
Observatory uses for quality control, archive production, and specialized
science reduction.  The EDPS will be used to replace all data reduction systems
using different workflow specifications that are currently used at the European
Southern Observatory.  }

   \keywords{
           methods: data analysis, methods: numerical, virtual observatory tools, techniques: miscellaneous
   }

   \maketitle

   \begin{figure*} [ht] \begin{center} \begin{tabular}{c}
           \hspace{-0.1cm}\includegraphics[width=9cm]{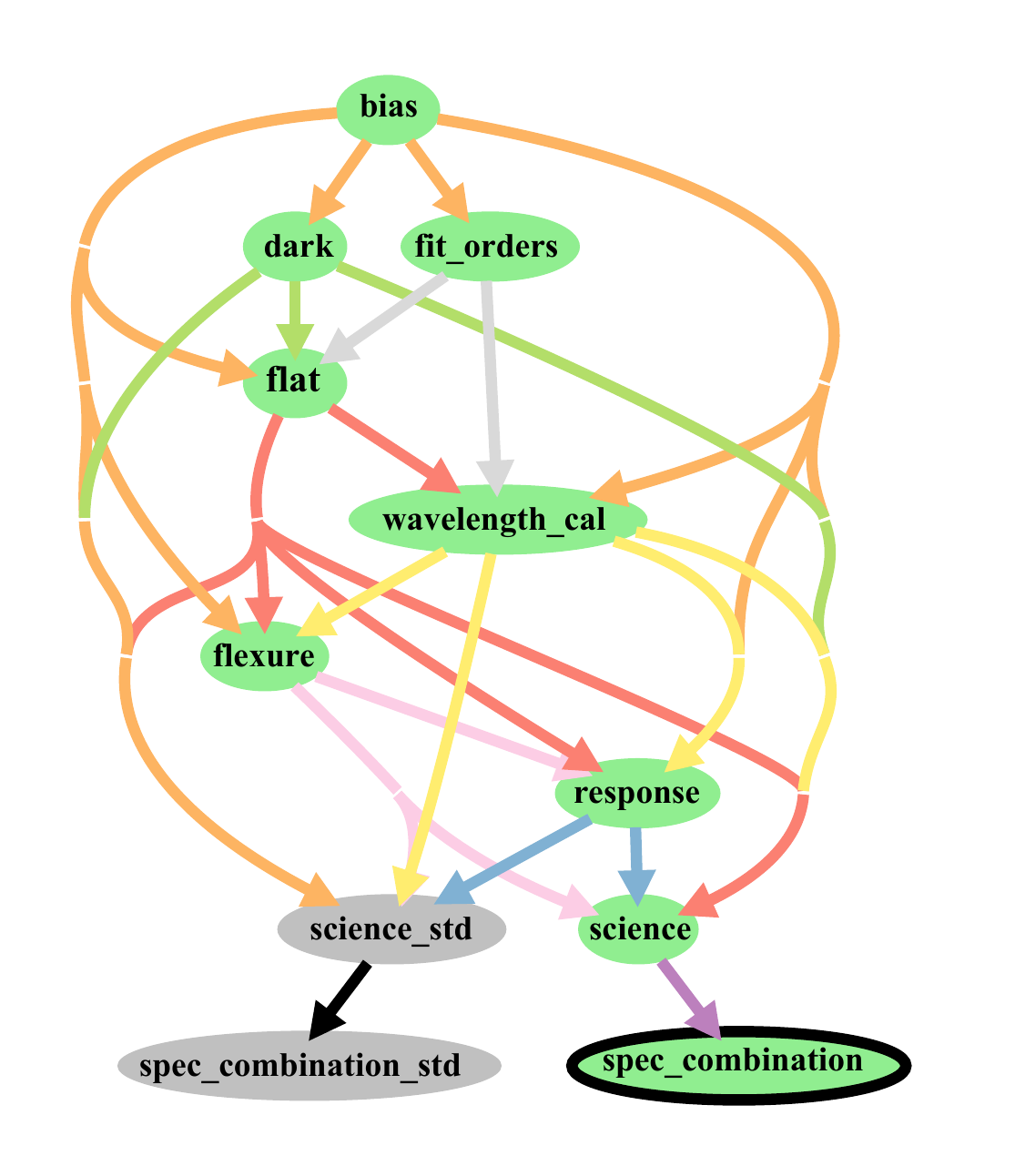}
           \hspace{0.1cm}\includegraphics[width=9cm]{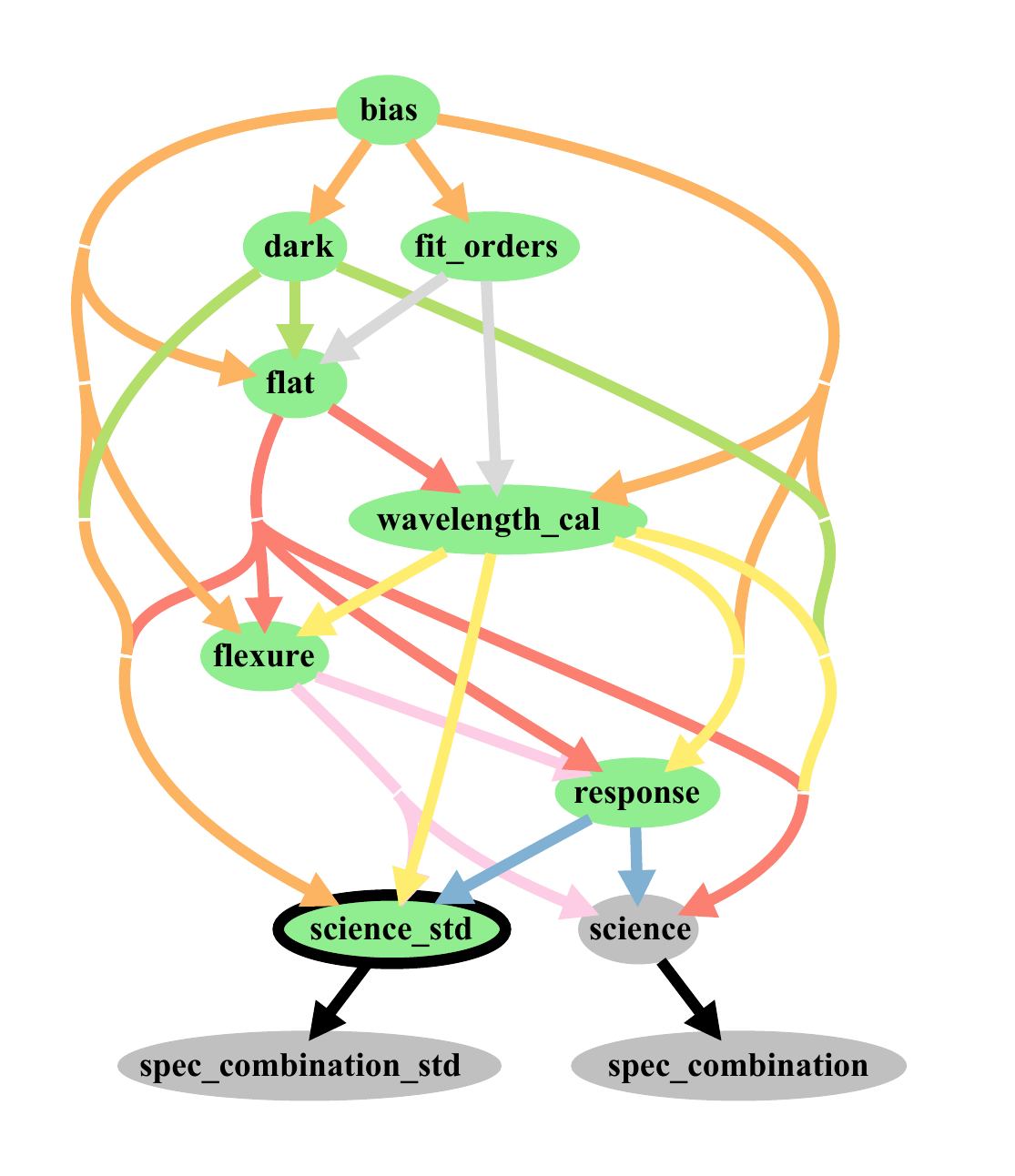}
   \end{tabular}\vspace{-0.8cm}
   \includegraphics[width=15cm]{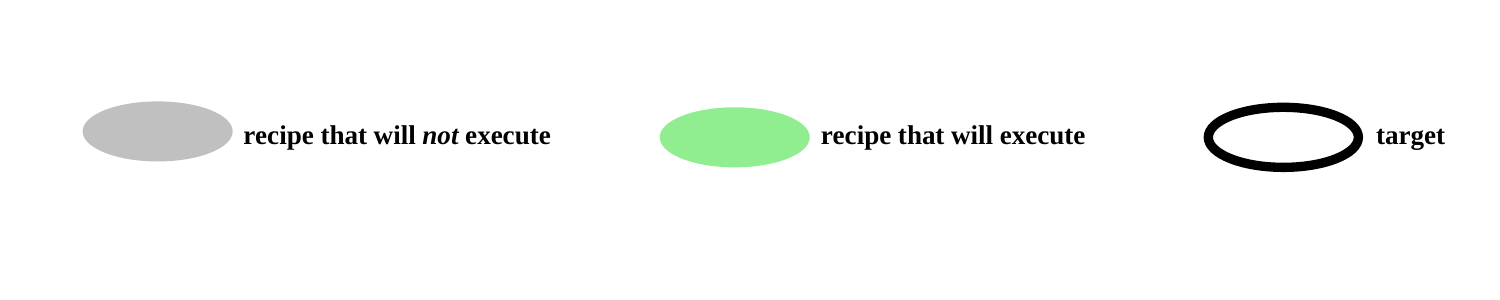}\vspace{-1cm}
   \end{center} \caption[example] { \label{fig:wkf_target} {Illustration of how
   the target determines the processing workflow. The two trees represent the
   same workflow specification when used with different targets.  Ellipses
   represent recipes, and the arrows represent output from one recipe which is
   fed to subsequent recipes. The  target recipe in each case is indicated by
   the black outline. The ellipses for recipes that are executed are coloured
   green, non-executed recipes are coloured grey. The corresponding file
   association step is shown in Fig.~\ref{fig:do}.  See the text for a detailed
   explanation. }

   } \end{figure*} 
   
   \keywords{data processing infrastructure, data reduction infrastructure,
   data reduction pipelines, data reduction workflows, data processing
   workflows, quality control, quality control pipelines, European Southern
   Observatory}

%
%

\section{Introduction}
\label{sec:intro}  

Modern astronomical telescopes and their instrumentation produce raw data that
record the scientific signal but also include the signature of the atmosphere,
telescope, instrument and detector(s). In addition, the raw data include noise
from different sources.  Data reduction is a necessary step before any data can
be scientifically analysed and interpreted. Observatories such as the European
Southern Observatory process data from their telescopes and distribute
processed data through open archives to their user communities
\citep[e.g.,][]{eso_archive,hubble_archive,keck_archive,alma_nrao_archive}.  At
the same time, the end users of the data are often in the best position to
carry out data reduction in a way that optimizes the quality of the reduction
for a particular purpose. 

In addition to the data reduction for scientific use, processing the raw data
also plays an important role for monitoring the quality of incoming data. A
rigorous quality control process can detect hardware and software issues, and
monitor the impact of the ambient conditions on the data. At ESO, observations
go through three levels of quality control. A first level is done right after
an observation is completed during the course of the night. This is followed by
two off-line quality control processes, one that focuses on the data quality of
one night and the characterization of the observing system performance, and
another one that investigates the long-term quality of science data.  In this
paper, we present the design of a data reduction system that automatically
adapts to such different use cases for data reduction and automatically derives
data reduction workflows for each of them.

%
%

\section{Pipelines and data processing workflows} 

\subsection{Basic concepts}

\label{sec:pipe}  

With the ever increasing complexity of astronomical instrumentation, it is
inevitable that the data processing is a multi-step process that requires using
highly specialized software tools. 
 
Observatories have long
recognized the need to provide data processing pipelines
\citep[e.g.,][]{muse_pipeline,gemini_pipeline,keck_pipeline,sdss_pipeline,nrao_pipeline}.
to support the exploitation of their data.
Such pipelines typically consist of a series of stand-alone modules ("recipes")
that have to be applied in sequence to the data. Each recipe is usually
designed to process a single data type as its {principal input. Associated
calibration files are used in the processing of the these input files. } The
algorithms used by the recipes usually have some parameters that can be set at
execution time.

The input files of a recipe can either be  unprocessed raw data,   the output
from previous processing steps, or pre-computed products called static
calibrations.  These recipes need to be combined into a data processing
workflow that specifies the sequence of processing steps to process a pool of
data.  Selecting the right input data and  executing individual steps in a data
reduction sequence is often carried out manually by the end user.  This process
requires significant expertise and effort. It is also error prone and raises
issues of reproducibility  and record keeping.  Automating data reduction
workflows is therefore highly desirable for individual science reduction
efforts, and mandatory for mass production.

A challenge for automating data reduction is that different data reduction
goals require different specific processing workflows.  For example, for
quality control a processing workflow might entail the processing of all
calibration data of a specific type. If the goal is instead to process science
data, only a subset of the calibrations may be needed. The processing workflow
in this case should select only the calibration data that are best suited for
the processing of the science, process them and then use them in combination
with the science data to create the final data product. This processing of
science data might be done in different ways, depending on their final usage.
For example, a processing workflow might skip some processing steps in order to
obtain higher signal-to-noise ratios at the expense of larger incidence of
artefacts.  Another example is that distortion correction by resampling of
imaging data are necessary when images are used for astrometry, but not desired
when the same images are used to search for weak lensing signals.  Different
processing workflows are also required for unsupervised processing of science
data for a science archive and interactive processing for a specific purpose.
The latter case often involves executing a recipe repeatedly, with different
values for parameters, to optimize the results before continuing with the next
workflow steps. 

To address these different data processing needs, ESO has long maintained
several different workflow systems. For science users, ESO Reflex
\citep{reflex_paper}  is offered as an interactive environment.  For each
supported pipeline, a pre-packaged data processing workflow is provided that
users can customize and execute on their own data. Reflex includes a host of
features for data organization, data  visualization and bookkeeping. In
addition to Reflex, ESO uses several other data reduction systems that run
workflows for different quality control levels and for archive processing.
These systems are only used within the ESO computing environment and cannot
easily be exported. ESO's different systems all run the same pipeline recipes,
but each is highly tuned towards running a specific kind of workflow.

\subsection{Unification of data processing workflows}

{It is clear that different workflows do not necessarily require different
systems to run them. A sufficiently flexible system could execute different
workflows that explicitly} encode the different data processing scenarios as
outlined above.  However, such explicit specifications would largely be
variants of the same underlying processing cascade. While there are substantial
differences between  the different workflows, their design is constrained by
the inter-dependencies of the recipes on each other. For example, if a recipe
is designed to process some flat-fields using preprocessed bias data, this
implies a fixed processing sequence.  The bias data have to be processed before
the flat recipe is executed.  This has to be taken into account in all
workflows.  In that sense, processing workflows for the different use cases
differ, but are also highly repetitive in their design. As a consequence, a
change in the underlying pipeline could require to change all workflows in a
very similar manner and a substantial effort would be needed to keep the
different processing workflows in sync. The fragility and huge effort in coding
and maintaining our huge collection of workflows motivated us to look for a
more robust and sustainable system for creating workflows. 

The goal of this paper is to present a system where all information necessary
to execute recipes is specified only once, and workflows for different
processing scenarios can then automatically be derived based on the available
data, configuration files and user input.  Such a system exploits the fact that
each reduction recipe requires specific data types as input and produces
specific data types as output.  The  recipes depend on each other in ways that
are defined by their input and output. These dependencies are therefore
identical for all data processing.  For example, a flat-field recipe will
always require raw flat-field frames and a combined bias frame as input, and
produce a combined flat-field frame as output.

The basic idea of the proposed system is to specify the dependencies of the
recipes that are in common to any possible workflow written for a given set of
recipes using a formal domain specific language. Individual processing
workflows are then derived by specifying a "target recipe"  to process a pool
of data. The target recipe is the recipe that produces the intended final
output product of a processing request. For example, if the target recipe is
the flat-fielding recipe, the system will derive the necessary processing
sequence that in this simple example consists of running first the bias
processing recipes followed by the flat-fielding recipe.  By contrast, if the
target is just the bias recipe, only that recipe is executed.

Fig.~\ref{fig:wkf_target} illustrates the concept of a target for the case of a
somewhat complex data reduction workflow, which is a simplified version of the
workflow for the  X-Shooter instrument on ESO's Very Large Telescope (VLT)
discussed in more detail in Sec~\ref{sec:xshooter}.  The recipes are shown as
ellipses, and the output of recipes that is fed to subsequent recipes are shown
as arrows. All  but the two recipes called  {\tt spec\_combination} and {\tt
spec\_combination\_std} also need raw data as  input which is not shown in the
figure. 

{The recipe that takes a raw spectral image of the science object
and produces a one dimensional extracted spectrum is called {\tt science}. A
subsequent recipe called {\tt spec\_combination} combines multiple extracted
spectra from the same target.  In addition, there are a number of calibration
recipes that include the usual {\tt bias}, {\tt dark}, and {\tt flat} recipes
that combine individual frames to produce calibration files to be applied to
the data by subsequent recipes.  Other calibration recipes are the {\tt
fit\_order} recipe that finds the location of spectral orders, the {\tt
wavelength\_cal} recipe that derives the wavelength solution from an arc lamp
image, and the {\tt flexure} recipe that computes the correction for instrument
flexure from a special calibration with pin holes. }

{A noteworthy feature of the workflow is the processing of the
standard star.  The {\tt response} recipe uses the raw data from a flux
standard star, and produces a response curve to be applied to the science
spectrum. In standard science processing, the spectrum of the standard star is
only used for that purpose. However, in some cases, the spectrum of the star is
also of scientific interest, and in those cases the spectrum should be fed to
the science recipe.  The response curve is then computed from a different
observation of a standard star.  The science recipe is therefore placed twice
on the workflow specification, once called {\tt science} and once called {\tt
science\_std}. Each of them is followed by a separate instance of the recipe to
combine spectra.  In that manner, each instance of the science recipe or the
combination recipe can be specified as a target separately.  The different
instances refer to the same recipes, but are fed with different input data. }

{This setup can be used to nicely illustrate the use of the target
concept. In standard science processing, the desired outcome is the combined
spectra of science observations. In this case, the target is the {\tt
spec\_combination} recipe that follows the science recipe. This is illustrated
in the left tree in Fig.~\ref{fig:wkf_target}. The standard star is only used
to create the response curve. The right tree in the figures shows the case
where the purpose of the processing is to produce a science spectrum of the
standard star.  Since standard stars are typically bright, a combination of the
individual extracted spectra is not always desired. In that case the {\tt
science\_std} recipe is chosen as  the target. Recipes that are executed in
each of the two scenarios can automatically be determined, and  are marked in
green in the figure.  }

\begin{figure} [ht] \begin{center} \begin{tabular}{c}
\includegraphics[width=9cm]{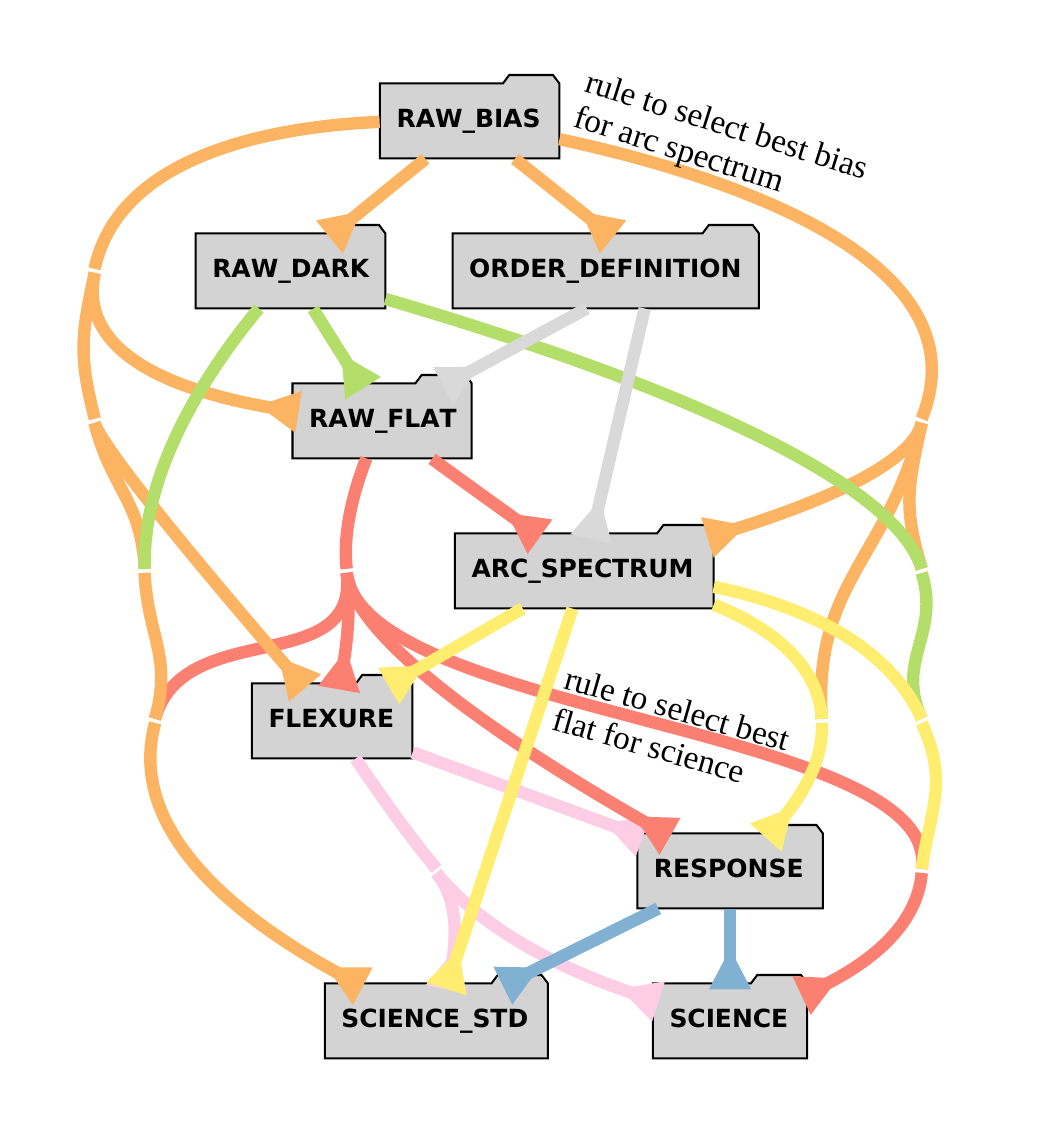} \end{tabular}
\end{center} \caption[example] { \label{fig:do} {Associations rules for the
workflows shown in Fig.~\ref{fig:wkf_target}. Each folder symbol represents a
set of raw data. Each line represents a rule for selecting the raw files at the
end of the line. The arrow at the start of the line indicates where the
information to select files is taken  from. The labels on  two of the lines
spell out the nature of those rules.} \vskip0cm  } \end{figure}

\begin{figure} [ht] \begin{center} \begin{tabular}{c}
\includegraphics[height=10cm]{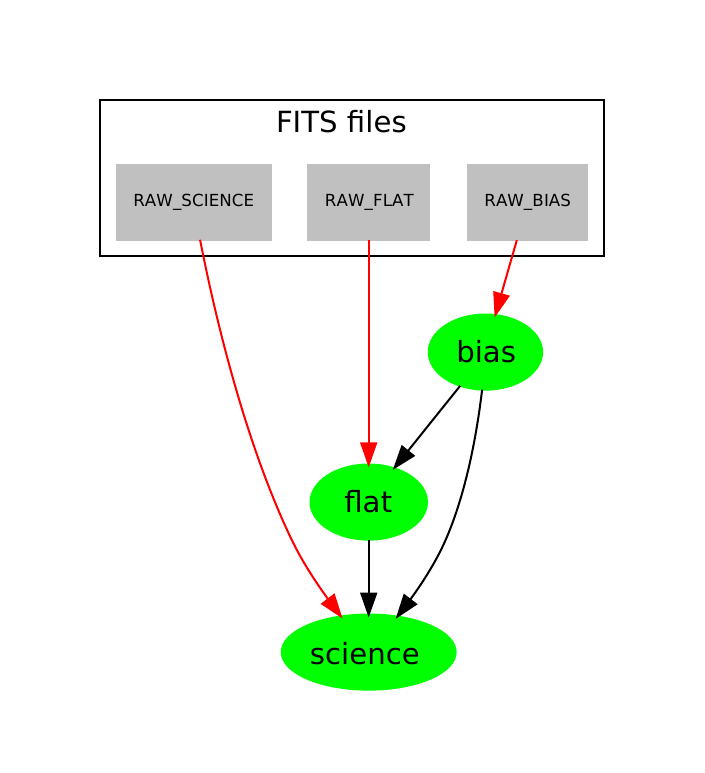} \end{tabular}
\end{center} \caption[example] { \label{fig:demo_workflow} EDPS visualization
of a simple workflow corresponding to the code listing in
Sec.~\ref{sec:edps_workflow}.  The green ovals are the tasks, grey squares the
data sources, red arrows the {\sl main inputs} and black arrows the associated
inputs of the tasks.} \end{figure} 

\section{Data classification and organisation}\label{sec:dataorganisation}

A data processing workflow by itself is not sufficient to carry out a full data
reduction. In order to do so, the correct data need to be identified and fed to
the individual recipes. The processing itself is therefore preceded by a step
that selects the data to be processed, and passes them to the right recipes.
Throughout this paper, we use the term ``data reduction workflow'' to include
both the data selection and subsequent data processing steps. 

Each of the recipes of the processing workflow as shown in
Fig.~\ref{fig:wkf_target} needs to be fed with a specific data type.  The data
type of a file can be derived from the data header based on some rules.  In the
simplest case, a single header keyword or a combination of them identify the
file type. The process of determining the file type of each available file is
called {\sl classification}.

Once all available files are classified,  a subset of them can be chosen for
processing. The process of selecting input files starts with the selection of
input for the target recipes, and then follows the processing cascade upstream.
The first step to organize the input for each recipe  is to group the
{principal  input files, i.e the input files the recipe is designed to
reduce. Hereafter, we refer to these files as the {\sl main input} of a recipe.
} For example, the {\tt spec\_combination} recipe shown in
Fig.~\ref{fig:wkf_target}  might process together all the spectra taken of the
same object. The processing of  the {\sl main input} requires additional input
files which have to be properly selected according to some criteria.  These
additional files are called {\sl associated} files, and the rules to select
them are called {\sl association rules}.  An example of this would be to select
a raw flat frame for the science processing based on the properties of the
spectra to be processed by the science recipes. {This is illustrated
in Fig.~\ref{fig:do}, which shows the tree of association rules that correspond
to the workflows of Fig.~\ref{fig:wkf_target}. Each folder in the figure
represents a set of input files, and the lines represent the rules to select
them. For example, there is a rule to select the best bias for the arc spectrum
which is identified in the figure. This rule needs as input some header
information from the {\tt ARC\_SPECTRUM} files, such as the time of observation
and the binning of the detector. The rule might state that the {\tt RAW\_BIAS}
must have the same binning as the {\tt ARC\_SPECTRUM}, and is as close as
possible in observing time. This rule is then applied to the data to select the
best match among all the provided {\tt RAW\_BIAS}es.}

It becomes clear that the description of data organisation is similar to  the
workflow itself, with the flow of information in the opposite direction. In
simple cases, the data reduction workflow can therefore be derived from the
description of the data organisation. However, as discussed in detail in
\cite{reflex_paper} and shown in Sec.~\ref{sec:xshooter}, in general this is
not possible. In that work, we opted therefore for an independent description
of the workflow and the data organisation and this approach was implemented in
the ESO Reflex software.  There are two disadvantages in this approach. First
of all, the two descriptions are at least a partial duplication of information.
In addition, while the descriptions are different, they have to be consistent
with each other.  Implementing and maintaining this consistency requires
significant manual effort. 

In this paper, we adopt a more efficient approach by first specifying the
complete data processing cascade, and then adding the data organisation
information, without any duplication, directly to the data processing tree from
which objective-specific workflows can be derived.

A specific implementation of this is discussed in Sec.~\ref{sec:edps}. Such a
design maintains the full flexibility to design complicated workflows as
discussed in \cite{reflex_paper}, while at the same time avoiding any
duplication in the description of the whole system.  The consistency between
the processing cascade and data organisation is inherent in the design.  The
description of the data processing cascade and data organisation rules can be
used to derive complete data reduction workflows, i.e.  the specific sequence
of data selection and data processing steps necessary to process a given set of
data and  a specified target.

%
%

\section{EDPS workflows}
\label{sec:edps}  

A general processing cascade system that implements the concepts discussed
above consists essentially of three parts. A description of the processing
cascade including the data organisation, an engine that parses the processing
cascade, derives the appropriate workflow and applies the rules to a given set
of input data to determine a sequence of recipes runs, and an executor that
schedules and executes the recipes.

Such a system, compatible with ESO's data reduction pipelines, has been implemented
as the {\sl ESO Data Reduction System (EDPS)}.  EDPS is entirely implemented in
the Python language \citep{python}, including the specification of processing
cascades.  EDPS allows the specification of processing cascades in a clear and
concise manner that describes the data reduction process without the need for
detailed knowledge of the implementation of other parts of EDPS or the Python
language.  In this section we describe the specification of processing cascades
in a level of details that illustrates the basic principles.  For a full
description of all feature on a more technical level we refer to
\cite{edps_manual}.

\subsection{Tasks} \label{sec:tasks} 

The central aspect of a processing cascade is the specification of the recipes
to be executed. This is typically the first step in writing a processing
cascade, because the recipes and their inputs and outputs are mostly known
already at the design stage of a pipeline.  In EDPS, the execution of a recipe
is specified in so-called "tasks".  Tasks are descriptions of processing steps
that can be executed independently once all their inputs are available. Most
tasks are used to execute recipes, but the concept of a task is more general
thus allowing other code to be executed as part of the processing cascade.  The
most basic properties of a task to be specified in a processing cascade are the
recipe to be executed and its input.  As noted in
Sec.~\ref{sec:dataorganisation}, each recipe has a {\sl main input} and
possibly one or more {\sl associated inputs} that are needed to process the
{\sl main input}. A fully functionally example code-snippet of a task
definition in EDPS is:

\begin{lstlisting}[language=Python]
flat_task = (task('FLAT')
             .with_recipe('flat_recipe')
             .with_main_input(raw_flat)
             .with_associated_input(bias_task)
             .build())
\end{lstlisting} 

Each task has a name for presentation purposes, in this example it is called
"FLAT".  At execution time, the target task of the execution can be specified
referring to the task by this name.  Tasks  have many optional parameters, some
of which are fairly complex to instantiate. Creating those objects using Python
constructors would require a long list of arguments.  To improve readability we
use the builder pattern syntax to provide values only for selected arguments.
These arguments can be recognized by their starting pattern {\tt.with\_}.  The
call of the .build() method indicates that the configuration is complete.

\subsection{Data Sources}\label{sec:data_sources}

The input of a task as shown in the above listing is either the output from a
preceding task, or some files that are available at the start of the execution
of a workflow. The former can be referred to by their object names, which is
"flat\_task" in the example above.  The latter have to be Flexible Image
Transport System (FITS) files \citep[][]{FITS}  that can be classified by
header keywords. In the processing cascade, these files have to be defined as a
"data source", that includes a {\sl classification} rule that specifies how to
recognize a file.  In its simplest form, it specifies one or several keywords
and their values to be used to define a data type. An example of such a simple
{\sl classification} rule is: 

\begin{lstlisting}[language=Python]
type_flat= classification_rule('rawflat', {'filetyp': 'RAW_FLAT'})
\end{lstlisting} 

that specifies that a file is reported as "rawflat" to the recipe if the header
keyword "filetyp" has the value "RAW\_FLAT". In general, much more complex {\sl
classification} rules can be specified, including support for alternatives and
conditional statements to analyse the headers of the files. Data from complex
instruments often require such more complicated rule.  The {\sl classification}
rules can then be used to define data sources, e.g.:

\begin{lstlisting}[language=Python]
raw_flat = (data_source('RAW_FLAT')
           .with_classification_rule(type_flat)
           .build())
\end{lstlisting} 

The above code-snippet specifies a data source named "RAW\_FLAT", and the
corresponding files are recognized by {\sl classification} rule given above.  A
full specification of data sources would also include a rule for how the files
should be grouped, i.e. how the  collections  of RAW\_FLATs that are fed to
each execution of a recipe are assembled.  EDPS also provides a syntax to
specify existing files as an alternative to the input created by a preceding
task.  This feature allows to run the same workflow using pre-processed input
files, e.g. using processed calibration files that are available in the ESO
Science Archive Facility \footnote{\url{https://archive.eso.org}} as an option when
downloading the data.

\subsection{Processing Cascades}\label{sec:edps_workflow}

In most cases, designing the basic processing cascade is the first task for the
development of a pipeline.  The EDPS syntax allows to specify and visualize
this overall architecture of a pipeline leaving out details such as the data
classification and selection.  This can be achieved by specifying the tasks
with their inputs and the corresponding data sources, using the syntax
described in the previous sections. An example of a processing cascade with
three tasks is shown in the following listing:

\begin{lstlisting}[language=Python]
# --------------- data sources ----------------

raw_bias = (data_source('RAW_BIAS')
             .build())

raw_flat = (data_source('RAW_FLAT')
             .build())

raw_science = (data_source('RAW_SCIENCE')
             .build())

# ------------------ tasks --------------------

bias_task = (task('bias')
             .with_main_input(raw_bias)
             .build())

flat_task = (task('flat')
             .with_main_input(raw_flat)
             .with_associated_input(bias_task)
             .build())

science_task = (task('science')
             .with_main_input(raw_science)
             .with_associated_input(bias_task)
             .with_associated_input(flat_task)
             .build())
\end{lstlisting} 

A visualization of this processing cascade, which can be created by EDPS, is
shown in Fig.~\ref{fig:demo_workflow}.  Further details can be added later,
turning the design into fully functional code for the execution of a multitude
of different workflows.  The simplicity and readability of the specification of
the basic processing cascade is an appealing property of the EDPS syntax.   

\subsection{Data Organization}\label{sec:edps_do}

As explained in Sec.~\ref{sec:dataorganisation}, a full data reduction
processing cascade needs additional specifications to allow the selection  of
{\sl associated input} that corresponds to the {\sl main input} of each task.
This selection is, in most cases, based on the closeness in time of the {\sl
associated} and {\sl main input} of a task, but usually also includes other
conditions, such as a match in detector binning or other data properties. The
way the data are associated is in most cases a property of the {\sl associated
input}, but in some cases different rules need to be applied depending on the
task that receives the input.  Association rules in EDPS can therefore either
be attached to the data source, or to the {\sl associated input} of a task. In
the simplest case, the syntax of these rules is similar to that of the {\sl
classification} rules shown above.  These associated calibrations are in some
cases optional, i.e. using them might improve the quality of a reduction but is
not strictly necessary. In other cases, the calibrations are needed only under
certain conditions. Such cases can be expressed in the rules and are taken into
account in the data organization.

The design of data organisation rules is an important consideration in the
design of a processing cascade, with potentially large impact on the overall
quality of the data reduction.  Taking optimal calibration data with
astronomical instruments is often expensive in terms of observing time.  A
trade-off between obtaining  more calibrations or investing more time in the
observations of the science object is often unavoidable. This then has
implications for the selection rules to be applied. An example of such a
situation is that calibration data ideally are from the same night. If an
instrument is sufficiently stable, then data from previous or following nights
up to some limit can be used for the calibration. Another example is that a
calibration product can be replaced with a different inferior data product.
The EDPS processing cascade data organisation can capture such  situations by
attaching several association rules to a data product. Each of the rules can be
assigned a quality code, and the achieved quality of the data selection is
reported at run time for each data set.

\subsection{Advanced features}\label{sec:advanced}

The features in the language to describe EDPS processing cascades are an
implementation of the general concepts described in Sec.~\ref{sec:pipe}. With
them, a wide variety of data reduction processing cascades and workflows can be
described, allowing EDPS to support the pipelines of all of ESO's operational
instruments, as well as many now-retired instruments.  EDPS also supports a
number of advanced features that can be used to implement more complex
scenarios. 

All the data organisation rules based on FITS header keywords discussed in the
previous section can also be formulated directly as Python functions, bypassing
the provided custom classes. This allows the implementation of complex logic
that acts on combinations of keywords. This feature is for example useful if a
pipeline supports multiple instruments or instrument modes, each of them
creating data with different header structures.

Another scenario that frequently occurs is that recipe parameters have to be
set differently for input data with different properties. An example of this is
that if a dedicated sky observation is available, the parameters for estimating
the sky levels are set differently than in the case the sky level has to be
derived from the science exposure itself.  A related scenario is that the
processing cascade itself depends on properties of input data, i.e. different
tasks are executed for data with different properties. An example of this is
that a recipe to combine data is only triggered if a minimum number of
exposures are available.  EDPS supports such data driven processing cascades.
Also in this case, complex logical functions can be used in the specification
of the parameters. A related feature is to allow the specification of
conditions for data sources to be used. 

EDPS processing cascades can also have parameters that provide the user with
choices on the data reduction workflow. One use case of workflow
parameters is to set parameters for several recipes simultaneously in cases
when these recipes parameters need to be  set consistently. Another use case is
to provide users with choices in the reduction flow, see
Sec.~\ref{sec:xshooter} for an example.

One supported feature that does not add any functionality, but improves the
readability of EDPS processing cascades is the possibility to group tasks into
subworkflows. A subworkflow is itself a processing cascade with several tasks,
and can be used within the main processing cascade similar to a task. A
subworkflow can also include other subworkflows in an hierarchical manner. 

\subsection{Example X-Shooter processing cascade}\label{sec:xshooter}

\begin{figure*} [hp] \begin{center} \begin{tabular}{c}
        \includegraphics[width=15cm]{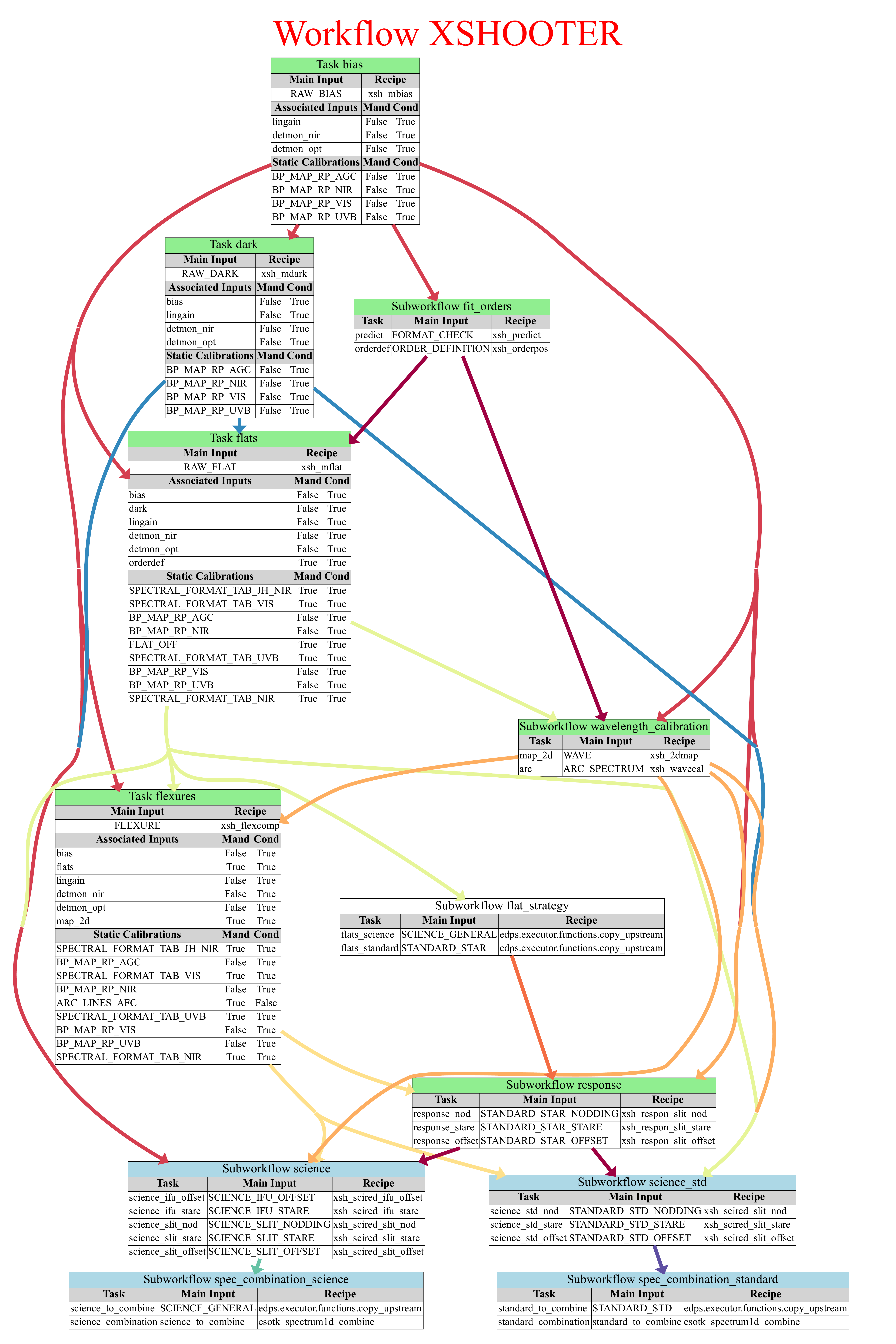}
\end{tabular} \end{center} \caption[example] { \label{fig:xshooter_workflow}
{EDPS visualization of the full X-Shooter processing cascade
specification. The topology of the processing cascade is essentially the same
is in Fig.~\ref{fig:wkf_target}, with one extra task called "flat\_strategy".
Each step of Fig.~\ref{fig:wkf_target} is now expanded with the full
information The individual steps of Fig.~\ref{fig:wkf_target} are either a
single task or a subworkflow with several task, as indicated  in the header of
each box. The colour coding of the header distinguishes between calibration
steps that produce calibration products (green) and science steps that could
serve as science target (blue). The subworkflow flat\_strategy does not produce
any products itself and only serves to redirect products produced by other
steps. }} \end{figure*}

{To illustrate the combination of  concepts discussed in the
previous sections, we show in Fig.~\ref{fig:xshooter_workflow} how they are
used for the pipeline of the complex X-Shooter instrument \citep{xshooter}.
The topology and basic steps of this processing cascade are identical to the
ones shown in Fig.~\ref{fig:wkf_target}, but this figure uses a representation
of a processing cascade which shows addition information for each task and
subworkflow.  EDPS can create such a graph alternatively to the previously
shown graph in Fig.~\ref{fig:demo_workflow}.  Each box represents the task or
subworkflow named in the coloured header for that box.}

{The X-Shooter instrument has 3 different arms, and can be used in 3
different observing modes, namely "stare", "nod" and "offset".  While the
overall data flow for the reduction of data taken with the different modes is
similar, they use different recipes for many of the steps. Some of the steps
are only used for one or two of the arms.   In total, the data reduction for
the X-Shooter instrument uses 18 recipes that take 65 different types of files
as input.  Despite this complexity, using subworkflows, the main processing
cascade can be displayed in a compact manner that is easy to understand by
grouping recipes that do similar steps for different modes into subworkflows. }

{Each task and subworkflow box in Fig.\ref{fig:xshooter_workflow} lists the
recipe and its {\sl main inputs} for each step.  A task includes only one
recipe, whereas a subworkflow might include several tasks and therefore several
recipes.  For each task and subworkflow, the different associated inputs are
listed, which correspond to the coloured lines that go into that step. Finally,
static calibrations are listed. Static calibrations are files such as line
lists that have a long-time validity and may not even be directly acquired at
the telescope or by the instrument.  For each associated input and static
calibration, it is shown whether they are mandatory input ('Mand') and whether
they are conditional ('Cond') (cf. Sec.~\ref{sec:data_sources}). Many of the
input files in the X-Shooter processing cascade are conditional because they
are only used for some of the arms. For example, the NIR arm does not require
bias or dark frames as input. }

{This processing cascade is also an example of a case where the data
reduction workflow cannot be derived from the description of the data
organisation alone. The reason for this is that in some cases the optimal flat
for the standard star is the one used by the science frame so that any
flat-fielding inaccuracies cancel out \citep[cf.][]{reflex_paper}.  The general
rule of "use a flat as close as possible in time to the file to process"
therefore does not apply in that case. This is implemented as the subworkflow
{\tt flat\_strategy} that does not include any regular recipe but only
conditional copy commands. A workflow parameter controls which strategy is used
to associate the flat frames, either the conventional "close in time" rule or
the strategy just described. }

\section{Execution of workflows}\label{sec:execution}

\subsection{Software architecture}

The EDPS processing cascades as discussed in the previous section can be coded
in Python without a detailed knowledge of the inner workings of EDPS.  The core
functionalities of EDPS are  to apply the data organisation rules to the
available data, derive the data processing workflow based on the processing
cascade description and the specified target, and schedule the execution of
recipes.

{The EDPS software has been implemented with a client-server
architecture \citep{edps_manual}. The server part is a persistent process that
carries out all the necessary processing based on the processing cascade
specification and the input data. The server program does not terminate after
the processing is completed but keeps listening for requests to process data.
The advantage of this approach is that the start-up tasks carried out by the
server need to be executed only once, and data that might be used in subsequent
processing are kept in memory. This improves the overall responsiveness of EDPS
after the initial start-up. }

{The server acts on requests that are submitted via a REST API as
used by standard web service clients. Such requests can in principle be
submitted via a web browser, but an EDPS client is provided as a more
convenient interface.  Requests include the location of the data, the
processing cascade specification, at least one  target and the value of any
recipe or workflow parameters that should be set to a value different from the
default.  After receiving a request, individual processing jobs are created and
executed by the EDPS server.  For a future release, we anticipate the
implementation of a dedicated GUI (Graphical User Interface) for more
convenient interactive processing. }

\subsection{Processing jobs}

The complete processing cascade description as discussed in the previous
section allows EDPS to process a given set of data with the specified pipeline.
The input data pool is typically a directory on a local disk, or a list of
files. 

The kind of processing carried out on these crucially depends on the specified
target in a processing request.   We want to re-iterate the importance of the
central concept of the target. If the target of the workflow shown in
Fig.~\ref{fig:demo_workflow} is the SCIENCE task, all files that match the
rules for the data source RAW\_SCIENCE will be processed.  Files of type
RAW\_FLAT  will only be processed if they are needed to process the selected
RAW\_SCIENCE files. Files of type RAW\_BIAS are processed if they are needed
either for the processing of the RAW\_FLATs or the RAW\_SCIENCEs.  By contrast,
if for example the BIAS task is specified as target, all files of type
RAW\_BIAS are processed irrespective of whether they are needed by any other
task.

When the EDPS request is submitted by the client, the server classifies all the
input files and organizes them according to the encoded rules in order to
enable the processing of the {\sl main input} of the target task. All the
information needed to execute each individual data processing step is collected
in a so-called "job". Jobs are individual processing units that can be executed
once all input data for that job are available.  The information contained in a
job includes the recipe to execute,  the {\sl main} and any {\sl associated
inputs}, the recipe parameters, and general attributes such as a task name, and
the completeness of the input files.  A single task can generate multiple jobs
from a request, depending on how many groups of files need to be processed.
For example,  if a request includes two science files and associated
calibration files and specifies the SCIENCE task of the processing cascade
depicted in Fig. \ref{fig:demo_workflow} as the target, then the following jobs
will be generated:

\begin{enumerate}

    \item Two jobs from the SCIENCE task, one for each of the RAW\_SCIENCE
            files to be processed.

    \item One job from the FLAT task for each group of RAW\_FLATS files to be
            processed. If the two RAW\_SCIENCE requires two different sets of
                RAW\_FLATS, then 2 jobs are generated.

    \item One job from the BIAS task for each group of RAW\_BIAS to be
            processed. Depending whether the biases for the different
                RAW\_FLATS and RAW\_SCIENCE are the same or not, the BIAS task
                could thus generate up to 4 jobs.

\end{enumerate}

\medskip 

If all the mandatory input data for a job are available, or can be produced by
the execution of other jobs, a job is marked as "complete". If this condition
is not met, a job is "incomplete". If a job is incomplete, the job and all of
its child jobs cannot be executed (cf. Sec. \ref{sec:edps_scheduling}).

\subsection{Job execution and scheduling}\label{sec:edps_scheduling}

The jobs created by EDPS then need to be scheduled for execution.  An obvious
constraint on the scheduling is that a job cannot be executed before all of its
input are available. If the input is the output of other jobs, they have to
be scheduled first.  EDPS will also detect if a recipe has been executed with
the same set of parameters and input files before, and in such a case re-use
the output produced by the previous execution.  This is commonly referred to as
"smart re-runs". 

These constraints by themselves do not imply a unique sequence of scheduling.
Hardware resources and the general objectives of the scheduling have to be
considered in order to choose a strategy for scheduling jobs.  If a large
amount of data is to be processed in batch mode, the objective usually is to
minimize the total processing time, given a set of hardware resources. However,
the situation is different in interactive mode where a user wants the jobs to
be executed in a sequence that allows interaction in a manner that they can
easily follow and understand the consequences of their interactions.  A typical
usage scenario for the interactive case is that the goal is to process several
independent sets of files with the target recipe. An example of this is to
process several sets of images. Each set includes images that overlap, but the
individual sets are spatially separated from each other. An image co-adding
recipe accepts one full set of overlapping images. Following
\cite{reflex_paper},  EDPS uses the concept of datasets, which are the files
that are processed together by the target recipes, together with all
calibration files that are needed to make the execution of the target recipe
possible.  An interactive user will then want to process one dataset at a time,
inspecting the intermediate data products  as the processing proceeds. The
processing is paused after the execution of each recipe to allow the review of
the products and possible re-execution of this step with different recipe
parameters. During such an interactive processing,  jobs that process data for
other datasets  should only be executed  if they do not slow down the
processing of the currently considered dataset.  Typically, processing one
dataset at a time will increase the overall processing time for all datasets.
The benefit of this sequence is the shortened waiting time for the user to
provide the interactive input for the first dataset. 

{EDPS supports three algorithms to define the order of processing {\em
before the start of any job execution}, and one that schedules the jobs
dynamically.  The static algorithms are  the "depth-first" (DFS),
"breadth-first" (BFS) and "type search" algorithms. The DFS scheduling covers
the interactive processing use case described above.  The BFS schedules jobs
independent of datasets or types, taking into account the number of jobs that
can be simultaneously submitted for execution. The type search scheduling
schedules all jobs of a certain input type in sequence.  The {\em dynamic}
scheduling means that the order is determined after the start of job
executions.  When a job is completed, all jobs in the waiting queue are
considered for execution.  Depending on the total number of cores and CPUs made
available to the server and the number of threads used by each recipe, jobs
that make the best use of the available resources are started. }

\subsection{Testing of processing cascades}

{EDPS processing cascades with many tasks, data sources and associated data
organisation are sufficiently complex to warrant extensive testing when changes
are made.  EDPS processing cascades are therefore accompanied by a
specification of data files  that can be used to test the processing cascades.
The corresponding FITS files are then created when the test is executed.  The
test itself consists of using the data organisation of the processing cascade
to produce the input for and trigger the tasks, without actually executing the
recipes themselves. The tests therefore verify the grouping  and  association
of data as well as the  structure of the processing cascades, thereby
identifying the most common errors in processing cascades.  The test
specifications are coded in JavaScript Object Notation (JSON) files that
describe the input files with their header keywords, the target tasks, and the
values of any processing cascade parameters.  In addition, the test
specifications include the expected jobs and their input files.  The EDPS test
facility runs the processing cascade specification on the automatically created
test files and compares the list of tasks that have been triggered and their
inputs with the expected outcome as specified in the test itself.  A test is
passed if all predicted tasks are triggered with their predicted inputs and no
unexpected tasks are triggered.  These tests can, for example, be automatically
executed in an integrated development environment (IDE), and they can also be
run automatically by a continuous integration system any time a change to a
processing cascade is made to check for any regression.}


\section{Summary}

We propose a data reduction system that automatically creates data reduction
workflows from  a processing cascade specifications. The necessary
specifications can be written following the typical development cycle of a
data reduction pipeline, where the reduction steps are specified first and the data
organisation information is added later as needed. 

Such a system avoids duplication of specification in two different ways: 1) All
information to derive workflows for different reduction use cases is encoded in
a single description (a so-called processing cascade) in an efficient way.  2)
The specification of the data organisation is fully integrated into the
processing cascade definition. This is different from previous systems, where
the processing workflow follows the data organisation which has to be specified
first and is therefore limited in its flexibility, or the approach advocated in
\cite{reflex_paper} where the data organisation is specified completely
independent of the processing workflow.

EDPS is an implementation of such a system. It was designed to run ESO's data
reduction pipelines, but workflows can be written for any pipeline that follows
the basic principle of stand-alone recipes that are executed in sequence, use
FITS files for input and output, and can be configured using parameter files.
{EDPS is in routine operation at ESO since April 2023, and has been
publicly released in October 2023
\footnote{\url{https://www.eso.org/sci/software/edps.html}}.  }
This release web page includes code, documentation and installation
instructions. 

\bigskip

\subsection*{Acknowledgments} The EDPS team at ESO includes Lodovico Coccato,
Wolfram Freudling, Enrique Garcia, Andrea Modigliani, Ahmed Mubashir Khan, Ralf
Palsa, Stanislaw Podgorski, Martino Romaniello, and Stefano Zampieri.

\bigskip

\bibliography{edps} 

\begin{thebibliography}{14}
\expandafter\ifx\csname natexlab\endcsname\relax\def\natexlab#1{#1}\fi

\bibitem[{{Brodheim} {et~al.}(2022){Brodheim}, {O'Meara}, {Mader}, {Berriman},
  {Brown}, {Fuhrman}, {Tucker}, {Gelino}, {Lynn}, \& {Swain}}]{keck_archive}
{Brodheim}, M.~N., {O'Meara}, J.~M., {Mader}, J.~A., {et~al.} 2022, in Society
  of Photo-Optical Instrumentation Engineers (SPIE) Conference Series, Vol.
  12186, Observatory Operations: Strategies, Processes, and Systems IX, ed.
  D.~S. {Adler}, R.~L. {Seaman}, \& C.~R. {Benn}, 121860H

\bibitem[{{Freudling} {et~al.}(2013){Freudling}, {Romaniello}, {Bramich},
  {Ballester}, {Forchi}, {Garc{\'\i}a-Dabl{\'o}}, {Moehler}, \&
  {Neeser}}]{reflex_paper}
{Freudling}, W., {Romaniello}, M., {Bramich}, D.~M., {et~al.} 2013, \aap, 559,
  A96

\bibitem[{{Lacy}(2016)}]{alma_nrao_archive}
{Lacy}, M. 2016, in Society of Photo-Optical Instrumentation Engineers (SPIE)
  Conference Series, Vol. 9910, Observatory Operations: Strategies, Processes,
  and Systems VI, ed. A.~B. {Peck}, R.~L. {Seaman}, \& C.~R. {Benn}, 991007

\bibitem[{{Lacy} {et~al.}(2015){Lacy}, {An}, {Benson}, {Arora}, {Fox},
  {Griffith}, {Kern}, {Lively}, {Lyons}, {Masters}, {Plank}, {Spolaor}, \&
  {Tody}}]{nrao_pipeline}
{Lacy}, M., {An}, R., {Benson}, J., {et~al.} 2015, in Astronomical Society of
  the Pacific Conference Series, Vol. 495, Astronomical Data Analysis Software
  an Systems XXIV (ADASS XXIV), ed. A.~R. {Taylor} \& E.~{Rosolowsky}, 425

\bibitem[{{Law} {et~al.}(2016){Law}, {Cherinka}, {Yan}, {Andrews}, {Bershady},
  {Bizyaev}, {Blanc}, {Blanton}, {Bolton}, {Brownstein}, {Bundy}, {Chen},
  {Drory}, {D'Souza}, {Fu}, {Jones}, {Kauffmann}, {MacDonald}, {Masters},
  {Newman}, {Parejko}, {S{\'a}nchez-Gallego}, {S{\'a}nchez}, {Schlegel},
  {Thomas}, {Wake}, {Weijmans}, {Westfall}, \& {Zhang}}]{sdss_pipeline}
{Law}, D.~R., {Cherinka}, B., {Yan}, R., {et~al.} 2016, \aj, 152, 83

\bibitem[{{Lemoine-Busserolle} {et~al.}(2019){Lemoine-Busserolle}, {Comeau},
  {Kielty}, {Klemmer}, \& {Schwamb}}]{gemini_pipeline}
{Lemoine-Busserolle}, M., {Comeau}, N., {Kielty}, C., {Klemmer}, K., \&
  {Schwamb}, M.~E. 2019, \aj, 158, 153

\bibitem[{{Lockhart} {et~al.}(2019){Lockhart}, {Do}, {Larkin}, {Boehle},
  {Campbell}, {Chappell}, {Chu}, {Ciurlo}, {Cosens}, {Fitzgerald}, {Ghez},
  {Lu}, {Lyke}, {Mieda}, {Rudy}, {Vayner}, {Walth}, \&
  {Wright}}]{keck_pipeline}
{Lockhart}, K.~E., {Do}, T., {Larkin}, J.~E., {et~al.} 2019, \aj, 157, 75

\bibitem[{{Romaniello} {et~al.}(2023){Romaniello}, {Arnaboldi}, {Barbieri},
  {Delmotte}, {Dobrzycki}, {Fourniol}, {Freudling}, {Grave}, {Mascetti},
  {Micol}, {Retzlaff}, {Rosse}, {Tax}, {Vuong}, {Hainaut}, {Rejkuba}, \&
  {Sterzik}}]{eso_archive}
{Romaniello}, M., {Arnaboldi}, M., {Barbieri}, M., {et~al.} 2023, The
  Messenger, 191, 29

\bibitem[{{Tran} {et~al.}(2020){Tran}, {Burger}, \& {Hack}}]{hubble_archive}
{Tran}, H.~D., {Burger}, M., \& {Hack}, W. 2020, in Astronomical Society of the
  Pacific Conference Series, Vol. 527, Astronomical Data Analysis Software and
  Systems XXIX, ed. R.~{Pizzo}, E.~R. {Deul}, J.~D. {Mol}, J.~{de Plaa}, \&
  H.~{Verkouter}, 587

\bibitem[{Van~Rossum \& Drake(2009)}]{python}
Van~Rossum, G. \& Drake, F.~L. 2009, Python 3 Reference Manual (Scotts Valley,
  CA: CreateSpace)

\bibitem[{{Vernet} {et~al.}(2011){Vernet}, {Dekker}, {D'Odorico}, {Kaper},
  {Kjaergaard}, {Hammer}, {Randich}, {Zerbi}, {Groot}, {Hjorth}, {Guinouard},
  {Navarro}, {Adolfse}, {Albers}, {Amans}, {Andersen}, {Andersen}, {Binetruy},
  {Bristow}, {Castillo}, {Chemla}, {Christensen}, {Conconi}, {Conzelmann},
  {Dam}, {de Caprio}, {de Ugarte Postigo}, {Delabre}, {di Marcantonio},
  {Downing}, {Elswijk}, {Finger}, {Fischer}, {Flores}, {Fran{\c{c}}ois},
  {Goldoni}, {Guglielmi}, {Haigron}, {Hanenburg}, {Hendriks}, {Horrobin},
  {Horville}, {Jessen}, {Kerber}, {Kern}, {Kiekebusch}, {Kleszcz}, {Klougart},
  {Kragt}, {Larsen}, {Lizon}, {Lucuix}, {Mainieri}, {Manuputy}, {Martayan},
  {Mason}, {Mazzoleni}, {Michaelsen}, {Modigliani}, {Moehler}, {M{\o}ller},
  {Norup S{\o}rensen}, {N{\o}rregaard}, {P{\'e}roux}, {Patat}, {Pena}, {Pragt},
  {Reinero}, {Rigal}, {Riva}, {Roelfsema}, {Royer}, {Sacco}, {Santin},
  {Schoenmaker}, {Spano}, {Sweers}, {Ter Horst}, {Tintori}, {Tromp}, {van
  Dael}, {van der Vliet}, {Venema}, {Vidali}, {Vinther}, {Vola}, {Winters},
  {Wistisen}, {Wulterkens}, \& {Zacchei}}]{xshooter}
{Vernet}, J., {Dekker}, H., {D'Odorico}, S., {et~al.} 2011, \aap, 536, A105

\bibitem[{{Weilbacher} {et~al.}(2020){Weilbacher}, {Palsa}, {Streicher},
  {Bacon}, {Urrutia}, {Wisotzki}, {Conseil}, {Husemann}, {Jarno}, {Kelz},
  {P{\'e}contal-Rousset}, {Richard}, {Roth}, {Selman}, \&
  {Vernet}}]{muse_pipeline}
{Weilbacher}, P.~M., {Palsa}, R., {Streicher}, O., {et~al.} 2020, \aap, 641,
  A28

\bibitem[{{Wells} {et~al.}(1981){Wells}, {Greisen}, \& {Harten}}]{FITS}
{Wells}, D.~C., {Greisen}, E.~W., \& {Harten}, R.~H. 1981, \aaps, 44, 363

\bibitem[{Zampieri {et~al.}(2024)}]{edps_manual}
Zampieri, S. {et~al.} 2024, EDPS Manual (Garching: ESO)

\end{thebibliography}
\bibliographystyle{aa} 

\end{document}